\documentclass[journal=langmuir,manuscript=article]{achemso} 
\usepackage[version=3]{mhchem} 

\usepackage{graphicx}  
\usepackage{bm}        
\usepackage{amssymb}   
\usepackage{color}
\usepackage{amsmath}
\usepackage{natbib}
\usepackage{units}
\setkeys{acs}{articletitle = true}

\usepackage{booktabs}

\author{\normalsize Wojciech Kwieci\'{n}ski}
\affiliation{Physics of Interfaces and Nanomaterials group, MESA+ Institute for Nanotechnology, University of Twente, Postbus 217, 7500 AE Enschede, The Netherlands}    
\email{w.kwiecinski@utwente.nl}
\author{Tim Segers}
\affiliation{Physics of Fluids group, MIRA Institute for Biomedical Technology and Technical Medicine, MESA+ Institute for Nanotechnology, University of Twente, Postbus 217, 7500 AE Enschede, The Netherlands}
\author{Sjoerd van der Werf}
\affiliation{Physics of Interfaces and Nanomaterials group, MESA+ Institute for Nanotechnology, University of Twente, Postbus 217, 7500 AE Enschede, The Netherlands}
\author{Arie van Houselt}
\affiliation{Physics of Interfaces and Nanomaterials group, MESA+ Institute for Nanotechnology, University of Twente, Postbus 217, 7500 AE Enschede, The Netherlands}
\author{Detlef Lohse}
\affiliation{Physics of Fluids group, MIRA Institute for Biomedical Technology and Technical Medicine, MESA+ Institute for Nanotechnology, University of Twente, Postbus 217, 7500 AE Enschede, The Netherlands}
\author{Harold J.W. Zandvliet}
\affiliation{Physics of Interfaces and Nanomaterials group, MESA+ Institute for Nanotechnology, University of Twente, Postbus 217, 7500 AE Enschede, The Netherlands}
\author{Stefan Kooij}
\affiliation{Physics of Interfaces and Nanomaterials group, MESA+ Institute for Nanotechnology, University of Twente, Postbus 217, 7500 AE Enschede, The Netherlands}

\title{\textbf{\Large Evaporation of dilute sodium dodecyl sulfate droplets on a hydrophobic substrate}}

\begin{document}

\section*{Abstract}

Evaporation of surfactant laden sessile droplets is omnipresent in nature and industrial applications such as inkjet printing. Soluble surfactants start to form micelles in an aqueous solution for surfactant concentrations exceeding the critical micelle concentration (CMC). Here, the evaporation of aqueous sodium dodecyl sulfate (SDS) sessile droplets on hydrophobic surfaces was experimentally investigated for SDS concentrations ranging from 0.025 to 1 CMC. In contrast to the constant contact angle of an evaporating sessile water droplet, we observed that, at the same surface the contact angle of an SDS laden droplet with a concentration below 0.5 CMC first decreases, then increases, and finally decreases resulting in a local contact angle minimum. Surprisingly, the minimum contact angle was found to be substantially lower than the static receding contact angle and it decreased with decreasing initial SDS concentration. Furthermore, the bulk SDS concentration at the local contact angle minimum was found to decrease with a decrease in the initial SDS concentration. 
The location of the observed contact angle minimum relative to the normalized evaporation time and its minimum value proved to be independent of both the relative humidity and the droplet volume and thus, of the total evaporation time. We discuss the observed contact angle dynamics in terms of the formation of a disordered layer of SDS molecules on the substrate at concentrations below 0.5 CMC. The present work underlines the complexity of the evaporation of sessile liquid-surfactant droplets and the influence of surfactant-substrate interactions on the evaporation process.

\section{Introduction}	
Surfactants constitute an enormous group of chemicals which are typically used to reduce the surface tension of liquids. Surfactants usually consist of one or more hydrophobic tails and a polar, hydrophilic head. They are commonly used in numerous industries such as food processing \cite{surf_food}, agriculture\cite{evap_leaf}, pharmaceutical \cite{surf_drug} and inkjet printing \cite{inkjet}. In many of these applications surfactant laden droplets are deposited on a substrate where they are subsequently left to evaporate \cite{evap_leaf, inkjet, graphene_coating}. There is a vast amount of literature on the evaporation and dissolution of single and multi-component sessile droplets \cite{ouzo, disolution_binary}. Furthermore, the influence of surface roughness \cite{pillars1, xpillars, pillars3, pillars4, roughness4} and its wettability \cite{evap_stripes, patterned1} on the evaporation process have been extensively investigated in both experimental and theoretical studies. Although the basic physical principles describing droplet evaporation for model cases have been established, many relevant areas remain relatively unexplored, one of which is the influence of surfactants on the evaporation process.

When a soluble surfactant is added to a liquid, its surface tension decreases with increasing surfactant concentration until a threshold value is reached, i.e. the critical micelle concentration (CMC), beyond which the surface tension is not affected any further \cite{physics_surfaces}. The decrease in surface tension caused by surfactants typically results in a lower contact angle. The lower contact angle results in a larger droplet perimeter and thus, in a longer contact line. The evaporation rate is maximal at the contact line singularity and therefore the addition of surfactants may lead to an increased evaporation rate.~\cite{popov} At the same time, insoluble surfactants can significantly reduce the evaporation rate by forming a resistive barrier against liquid molecule diffusion.~\cite{evap_block} Furthermore, a non-uniform surface concentration of surfactants  leads to a surface tension gradient, which induces Marangoni flow, in turn resulting in enhanced mixing inside an evaporating droplet.~\cite{surf_flow} Surfactant induced flows can also greatly alter the evaporation process of droplets loaded with colloidal particles.~\cite{evap_colloids}

Next to surfactant behavior in the liquid bulk \cite{Lindman_micelles}, at the liquid-gas \cite{evap_block, surf_flow} and liquid-liquid \cite{surf_liquid_surfaces} interfaces, the evaporation process is potentially further complicated by surfactant adsorption at the solid substrate, which can lead to dramatic changes in substrate wettability and therewith to phenomena, such as superspreading \cite{superspreading1, superspreading2} or auto-phobing \cite{autophoby, autophoby2}. The surfactant adsorption process, the conformation of the adsorbed molecules, and their effect on droplet evaporation are still poorly understood since they proved to be challenging to study. The main difficulty arises from the fact that the adsorbing surfactant layers are usually only a few nanometers in thickness and the adsorption process occurs in a liquid, which limits the available experimental techniques \cite{SurfOrganization, SDSadsorption, SDS_ads_hydrophylic}. Furthermore, surfactant adsorption is highly specific for the surfactant-substrate combination and their chemical and physical properties resulting in a huge parameter space. Dynamically changing conditions and a moving contact line, present in the case of an evaporating droplet are complicating the problem even further. 

For the extensively studied surfactant sodium dodecyl sulfate (SDS), an anionic surfactant commonly used in industry and research, adsorption at the solid-liquid interface was studied only for static conditions where the whole solid surface was submerged in the surfactant dispersion.~\cite{SDSadsorption, SDS_ads_hydrocarbon, SDS_ads_hydrophylic} It has been shown that the layer of SDS molecules adsorbed  to a hydrophobic surface has a maximum thickness for an SDS concentration of approximately 7 mM, thus, at a concentration below the CMC (8.2 mM).~ \cite{SDSadsorption, SDS_ads_hydrocarbon} Furthermore, not only the thickness of the surfactant layer changes with varying SDS concentration, but also its structure, i.e. from amorphous to micelle-like agglomerates \cite{SDSadsorption}. Even though the evaporation of SDS laden droplets has been studied in the past, the influence of surfactant adsorption at the solid-water interface on the evaporation process has not been investigated in  full detail \cite{SDS_diffusion_evap}. 

Here, we studied the evaporation of aqueous SDS laden sessile droplets  from a hydrophobic substrate at SDS concentrations ranging from 0.025 to 1 CMC. We aimed to investigate the potential effect of SDS adsorption and layer formation at the solid-liquid interface on the evaporation process. 

\section{Experimental Details}

\subsection{SDS solution preparation}

Sodium dodecyl sulfate (SDS) with a purity of $\geqslant99\%$ was purchased from Sigma-Aldrich and used without further purification. Water from a Milli-Q system (resistivity $=18.2~ \unit{M\Omega\cdot cm}$) was used for solution preparation. Based on literature, the critical micelle concentration (CMC) for an aqueous SDS solution of 8.2 mM was used.~\cite{SDSsurfacetension, sds_cmc} The dependence of the surface tension of SDS solutions on the surfactant concentration has been reported previously.~\cite{sds_cmc} All solutions were prepared in glass containers that were first cleaned with acetone, then with ethanol and finally rinsed with Milli-Q water. The solutions were used within 24 hours from their preparation. In the present work, the SDS concentrations are given relative to the CMC. The SDS concentrations mentioned in the description of the evaporation process refer to the initial SDS bulk concentration. The preparation accuracy of the surfactant solutions is estimated to be within 0.002~CMC.

\subsection{Substrate preparation}

Single side polished Si(100) wafers (Okmetic) were diced into 1.5x1.5 $\unit{cm^{2}}$ pieces and cleaned in an ultrasonic bath. First in acetone, and subsequently in ethanol, both for 10 minutes. Afterwards, the silicon substrates were dipped in   a freshly prepared Piranha solution (3:1 v/v mixture of sulfuric acid (96\%, Merck) and hydrogen peroxide (30\%, Merck)) for approximately 15 minutes and rinsed thoroughly with Milli-Q water thereafter. Subsequently, the substrates were transferred to a glass vacuum chamber where chemical vapour deposition (CVD) of PFDTS (1H,1H,2H,2H-perfluorodecyltrichlorosilane, 97\%, Abcr GmbH) was performed to hydrophobize the surface. CVD was realized by placing $100~\unit{\mu L}$ of PFDTS in a vacuum chamber, which was evacuated and PFDTS was left to evaporate for approximately 12 hours. This small amount of PFDTS was evaporating within a few minutes. Long reaction times ensured that all the chemical reactions were fully terminated. Later, the substrates were annealed in an oven at $100\unit{^{o}C}$ for 1 hour to enhance diffusion and the silanization process, to ultimately obtain a higher fraction of covalently bound silanes. After cooling down to room temperature, the substrates were submerged in chloroform and put in an ultrasonic bath for 15 minutes to remove excess PFDTS that was not chemically attached to the surface. As a result of this procedure we obtained smooth, hydrophobic surfaces with typical RMS roughness values of only $140~\unit{pm}$, as determined with an AFM (see AFM image in the Supporting Information).

Prior to every experiment, the substrates were cleaned in an ultrasonic bath, first in acetone for 10 minutes, and subsequently in a 50/50 (v/v) mixture of water and ethanol for another 10 minutes. After each cleaning step, the substrates were rinsed thoroughly with Milli-Q water and dried under a stream of nitrogen gas.

\subsection{Measurement set-up}

Droplet volumes smaller than $1~\unit{\mu L}$ were used to avoid the impact of gravitational forces on the shape of the droplets. Depending on the droplet size, two different deposition methods were used. For the $0.5~\unit{\mu L}$ droplets, a glass syringe (Hamilton, volume-100  $\unit{\mu L}$) with a cannula needle was used in combination with an automatic dispensing system of a contact angle measurement set-up (OCA15+, DataPhysics Instruments GmbH). The same deposition method was used for the receding contact angle measurements. To achieve droplet volumes smaller than 0.5 $\unit{\mu L}$, a single nozzle printhead (AD-K-501, Microdrop Technologies GmbH) with a nozzle diameter of $70~\unit{\mu m}$ was used. Droplets of approximately $0.5~\unit{nL}$ could be jetted at frequencies up to 100 Hz. By jetting several droplets onto each other, the droplet volume could be controlled over a range smaller than that achievable with the needle deposition method. The droplet volumes studied in this work were varied over two orders of magnitude ranging from 3.0 nl to 0.5 $\unit{\mu L}$. 
For both the needle and the printhead deposited droplets, the evaporation times were approximately two orders of magnitude longer than the time required to deposit the droplets. Both the needle and the pipette were thoroughly rinsed with Milli-Q water before and after the SDS solutions were introduced.

After droplet deposition, the evaporation process was recorded using a CCD camera (pco.pixelfly, PCO AG) and a lens with adjustable magnification (Navitar). The optical resolution of the images varied between 1 and 4 $\unit{\mu m/pixel}$ depending on the droplet size. The humidity in the measurement chamber was controlled using a home-made humidity control setup. The setup consisted of a mixing chamber where dry nitrogen gas can enter either directly (to lower the humidity) or through the water column (to increase the humidity). From the mixing chamber, gas was pumped at a low rate (5~\unit{L/h}) to the measurement chamber in such a way that it did not flow directly onto the evaporating droplet. The humidity was measured by a sensor placed in the measurement chamber close to the evaporating droplet and it was used to control the relative humidity at $44\pm2\unit{\%}$, if not stated otherwise. The temperature in the measurement chamber was fixed at $21\pm1\unit{^{o}C}$.

\subsection{Image processing}

The recorded images were processed using custom made Matlab (2017b version, The MathWorks, Inc.) scripts. In the analysis, the droplets were assumed to be axi-symmetric. All droplets were smaller than the capillary length, which allowed us to use the spherical cap approximation to determine the contact angle of the droplets by fitting a circle to the droplet's contour that was extracted by the script. The droplet volume was obtained by dividing the droplet into disks exactly one pixel in height and by summing up their volumes. The bulk SDS concentration inside the evaporating droplets was calculated based on the known value of the initial surfactant concentration, the initial droplet volume, and the droplet volume in each time frame extracted from the images. In some cases the evaporation time $t$ was normalized to the total evaporation time $T$ to aid data comparison, i.e. $\tilde{t}=t/T$. The measurements on the influence of the SDS concentration, the humidity, and the droplet size on the evaporation process were repeated at least four times of which the contact angle and radius evolutions were averaged.

\section{Results and discussion}

\begin{figure}[th] 
	\begin{center}          
		\includegraphics[width=0.5\columnwidth]{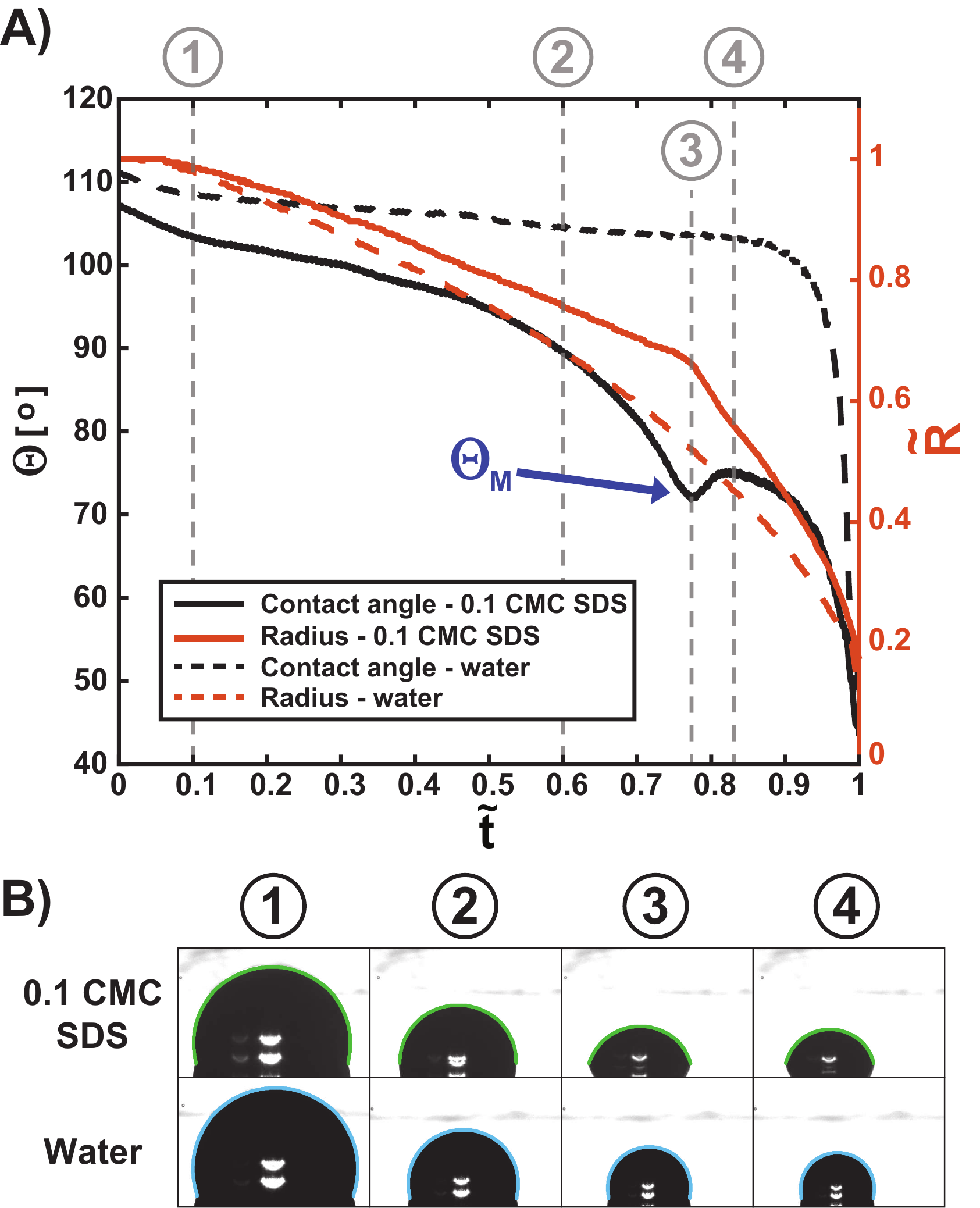} 
		\caption{A) Contact angle (black lines) and normalized radius (red lines) of an evaporating $0.5~\unit{\mu L}$ pure water droplet (dashed lines) and of an evaporating $0.5~\unit{\mu L}$ SDS laden droplet (0.1 CMC, solid line) as a function of the normalized time. B) Snapshots of the evaporating SDS laden droplet (top) and of the pure water droplet (bottom) at the different moments in time marked in panel A). The droplet contours, as found by the image processing script, are marked by the green line for the SDS laden droplet and in blue for the water droplet. Note that the apparent contour at the bottom edge of the images, which is not fitted, is part of the mirror image, i.e., an optical artefact.}
		\label{Fig1}
	\end{center}  
\end{figure}

Figure~\ref{Fig1} shows a comparison between the evaporation process of a pure water droplet and that of a SDS laden droplet with an initial bulk concentration of 0.1 CMC. The time is non-dimentionalized by the time $T$ it takes to evaporate the whole droplet, $\tilde{t} = t/T$. Note that the water droplet (dashed lines in Figure~\ref{Fig1}A) evaporates in the constant radius mode (CR) during approximately the first 5\% of the total evaporation time. Subsequently, it evaporates in the constant contact angle mode (CCA) throughout 90\% of the total evaporation process, switching to the mixed mode only during the last 5\% of its lifetime. For the various evaporation modes, we refer to refs.~\cite{evap_modes_1, evap_modes_4, evap_modes_5} and the review articles~\cite{evap_modes_2, evap_modes_3}. The SDS laden droplet (solid lines in Figure~\ref{Fig1}A) also evaporates in the CR mode during the first 5\% of the total evaporation time. However, after this initial stage, it switches to a complex mixed mode rather than to the CCA mode as in the case of a pure water droplet. Until approximately half of the total evaporation time, both the contact angle and the radius were observed to slowly decrease. Approximately halfway the evaporation process, the contact angle drops rapidly to reach its minimum value of $\Theta_{M}=72^\circ$ (point 3 in Figure~\ref{Fig1}A and B). Subsequently, the droplet radius starts to decrease rapidly and the contact angle increases to reach a local maximum at $\Theta=75^\circ$ (point 4 in Figure~\ref{Fig1}A). Thus, in contrast to a pure water droplet, the contact angle curve of the SDS laden droplet has a local contact angle minimum $\Theta_M$. 

\begin{figure}[th] 
	\begin{center}          
		\includegraphics[width=0.6\columnwidth]{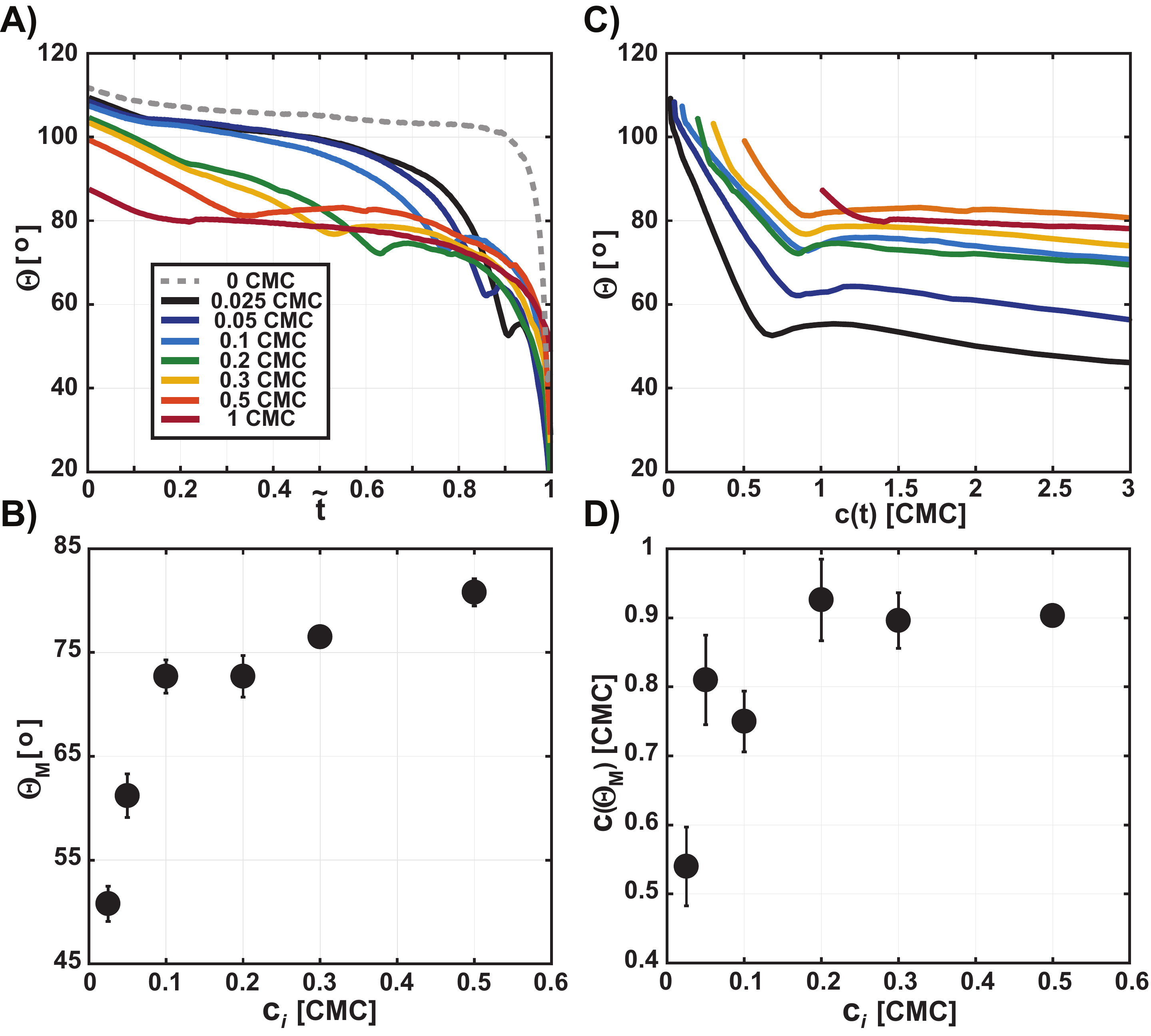} 
		\caption{A)~Contact angle averaged over 4~measurements of evaporating $0.5\unit{\mu L}$ SDS laden droplets with different initial SDS concentrations as a function of normalized time. B)~Contact angle at the local minimum $\Theta_M$ versus the initial SDS concentration. C)~Contact angle averaged over 4~measurements versus SDS bulk concentration during the evaporation of $0.5\unit{\mu L}$ SDS laden droplets. D)~Bulk SDS concentration at $\Theta_T$ versus initial SDS concentration. The error bars show standard errors of the mean.}
		\label{Fig2}
	\end{center}  
\end{figure}

The dependence of the local contact angle minimum on the SDS concentration is investigated in Figure~\ref{Fig2}A, i.e. it shows the contact angle evolution of evaporating droplets with initial SDS concentrations $c_i$ ranging from 0.025 CMC up to 1 CMC. A local contact angle minimum can be observed for every initial concentration, except for 1 CMC. Furthermore, several interesting trends can be observed with varying initial SDS concentration. First, as expected, the initial contact angle of the droplets decreased with increasing SDS concentration due to a decrease in surface tension. Second, it is observed that the lower the initial SDS concentration, the lower the contact angle at the local minimum $\Theta_{M}$ (Figure~\ref{Fig2}B). Especially for $c_{i}<0.1~\unit{CMC}$, the $\Theta$ decreases to values as much as 35$^\circ$ below those observed during the evaporation of a droplet with $c_{i}= 1 ~\unit{CMC}$. Finally, it is observed that the lower the initial SDS concentration, the later the local contact angle minimum occurs during the evaporation process, see Figure~\ref{Fig2}A.

When the contact angle of the droplet would only be governed by its liquid-air interfacial tension, one would expect a minimal contact angle for sessile droplets with $c_{i}\geqslant1~\unit{CMC}$, or, for droplets of which the bulk concentration reaches 1~CMC during evaporation. Thus, the observed local minima cannot be explained from the liquid-air interfacial tension alone. 
It may be expected that the occurrence of the local contact angle minimum is correlated to the instantaneous SDS concentration $c(t)$ in the evaporating droplet, see Figure~\ref{Fig2}C. Note that the bulk concentration at which $\Theta_M$ occurs decreases with a decrease in initial concentration to values  as much as 20-50\% below the CMC, see also Figure~\ref{Fig2}D. Also note that the local contact angle minimum \emph{always} occurs at a bulk concentration below 1~CMC. Thus, the local contact angle minimum cannot be explained by micelle formation in the liquid bulk and its onset cannot be correlated to a given SDS bulk concentration.

\begin{figure}[h] 
	\begin{center}          
		\includegraphics[width=1\columnwidth]{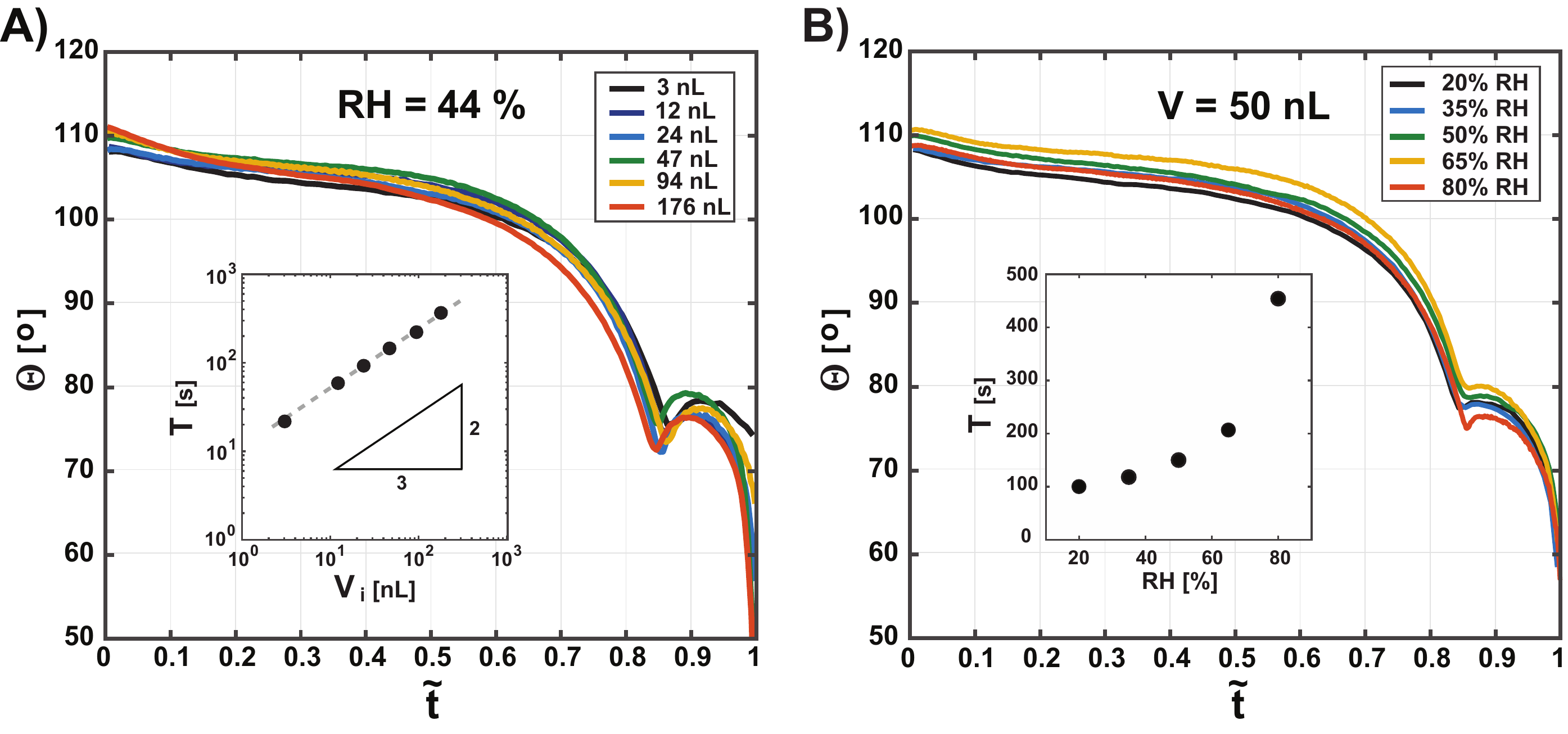} 
		\caption{ A) Contact angle averaged over 4~measurements versus normalized evaporating time for 0.05~CMC droplets of which the initial volume was varied. The relative humidity was $44\pm2\%$. The inset in panel A) shows the dependence of the total evaporation time $T$ on the initial volume of the droplet $V_{i}$. Note that $T$ scales as $V_i^{2/3}$. B) Contact angle curves averaged over 4~measurements as a function of the normalized evaporation time for 50~nL 0.05~CMC droplets for varying relative humidity (RH). The inset in panel B) shows that the total evaporation time was varied from approx. 100 to 500~s through the RH. }
		\label{Fig3}
	\end{center}  
\end{figure}

 To investigate whether the occurrence of the local contact angle minimum is dependent only on the bulk surfactant concentration inside the evaporating droplet, the initial droplet  volume and relative humidity were varied. Figure~\ref{Fig3}A shows the contact angle of evaporating 0.05 CMC droplets with different initial volumes (RH=44\%). The initial droplet volume $V_i$ was varied using the single nozzle printhead. Surprisingly, despite the fact that the initial volume was varied over two orders of magnitude (inset in Figure~\ref{Fig3}A) no significant differences or trends can be observed. The minor variations in the results can be attributed to small differences in the substrate condition or temperature fluctuations in the room. 
Furthermore, the total evaporation duration scales linearly with $V_{i}^{2/3}$ which is typical for diffusion driven evaporation \cite{evap_modes_1}.

The evaporation rate was varied through the relative humidity, see Figure~\ref{Fig3}B. Droplets with $c_{i}=0.05~\unit{CMC}$ and with a constant volume of 50~nL were used. Note that, varying the relative humidity in the measurement chamber did not influence the evaporation process, even though the total evaporation time $T$ varied by a factor of 5 (inset in Figure~\ref{Fig3}B). 

Figures~\ref{Fig3}A and B demonstrate the robustness of the local contact angle minimum during the evaporation of dilute SDS laden droplets, i.e. the phenomenon is independent of the droplet volume and the total evaporation time. This universality suggests that stochastic contact line pinning due to e.g. contaminant particles as the mechanism behind the local contact angle minimum can be excluded. Furthermore, it can be concluded that the timescale of the mechanism responsible for the local contact angle minimum is much faster than the time scale of evaporation, since no changes to the droplet evaporation behaviour were observed after varying the total evaporation time by almost two orders of magnitude.

\begin{figure}[h] 
	\begin{center}          
		\includegraphics[width=1\columnwidth]{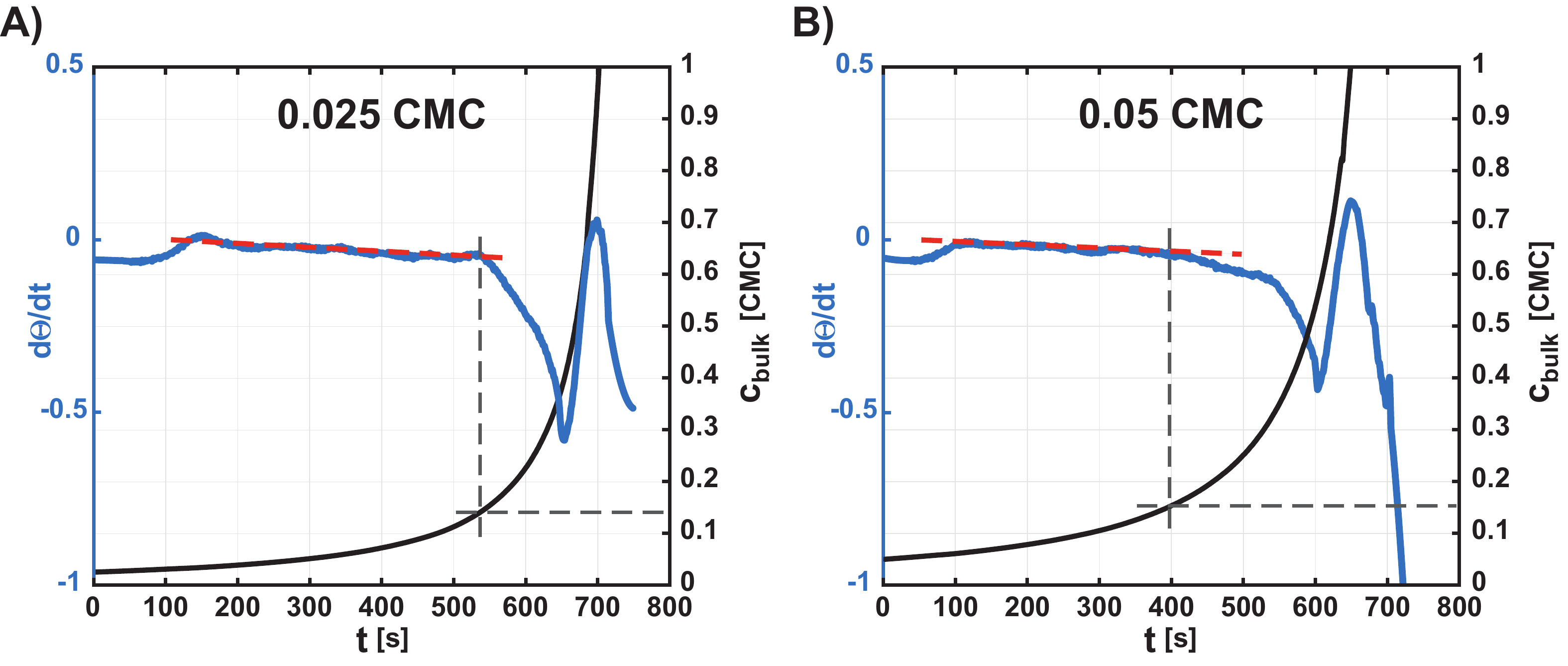} 
		\caption{Rate of change of the contact angle with respect to time $d \Theta / dt$ (blue lines) and SDS bulk concentration (black lines) as a function of time for evaporating 0.025~CMC (A) and 0.05~CMC (B) droplets. The dashed red lines show the average rate at which $d\Theta / dt$ changes during the first part of the evaporation process. The dashed gray lines mark the bulk SDS concentration at which the rate of change of $d\Theta /dt$ starts to deviate from that during the first part of the evaporation process. }
		\label{Fig4}
	\end{center}  
\end{figure}

The change in contact angle with the bulk SDS concentration is now investigated in more detail.  Figure~\ref{Fig4} shows the derivative of the contact angle $d\Theta / dt$ (left axis) and the bulk SDS concentration (right axis) as a function of time for 0.025~CMC and 0.05~CMC droplets. Note that for both cases the contact angle starts to decrease markedly faster when the bulk concentration amounts to approximately 0.15~CMC. This strongly suggests that the mechanism responsible for  the local contact angle local minimum sets in at bulk SDS concentrations as low as 0.15~CMC.  

\begin{figure}[h] 
	\begin{center}          
		\includegraphics[width=.6\columnwidth]{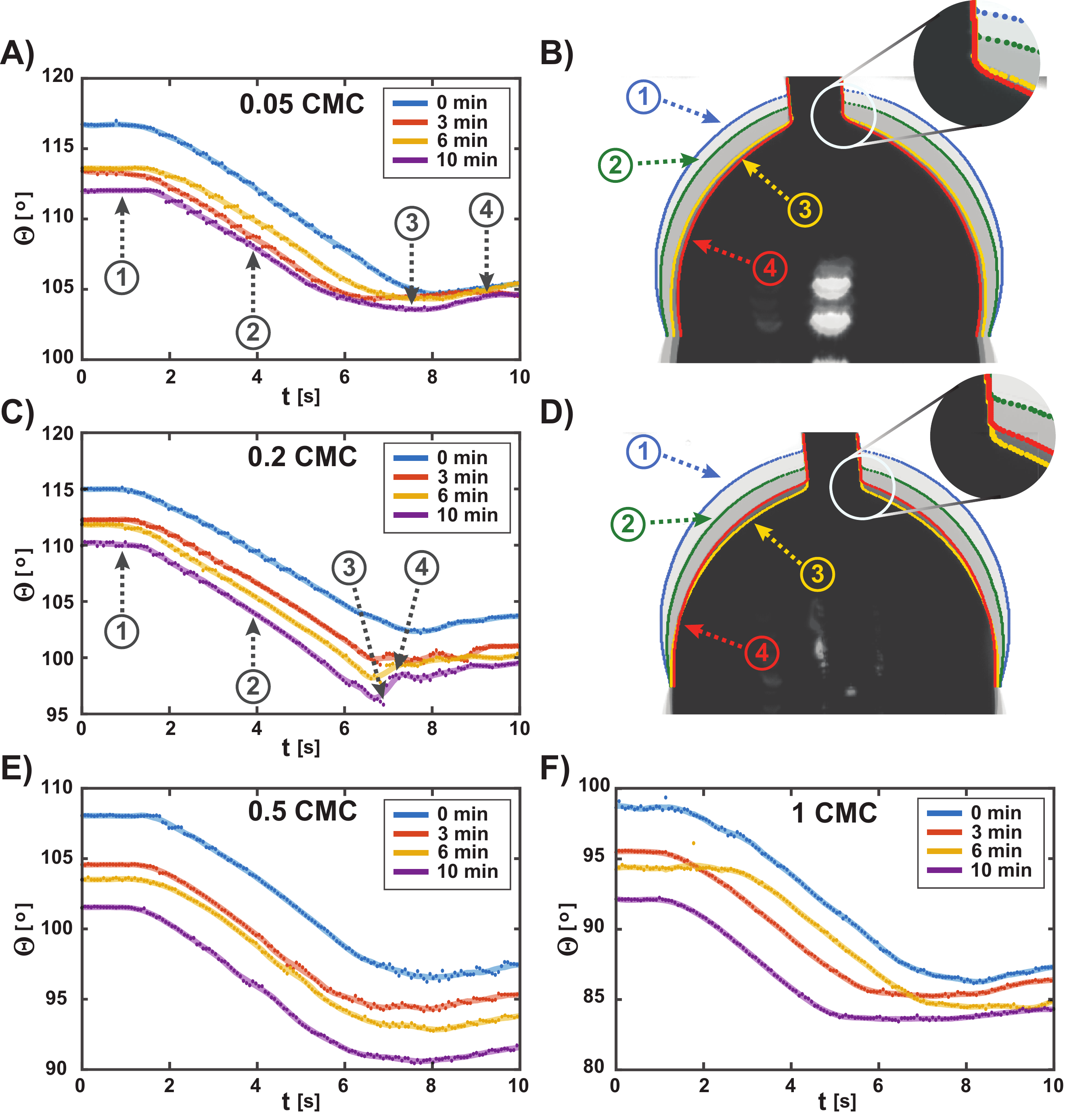} 
		\caption{A), C), E), F) Receding contact angle measurements for the SDS solutions with concentration of 0.05, 0.2, 0.5 and 1 CMC. Measurements were performed instantly after droplet deposition and later after 3, 6 and 10 minutes of waiting time. Contact line pinning, growing with time, visible as rapidly increasing contact angle, can be observed for 0.2 CMC solution. B) and D) snapshots of droplet during the measurement (after 10 minutes of waiting time) for 0.05 and 0.2 CMC solutions in different time points marked on panels A) and C). Inset on panel C) clearly shows droplet jump caused by the contact line pinning which is not present on inset on panel A).}
		\label{Fig5}
	\end{center}  
\end{figure}

To investigate whether SDS concentrations around 0.15~CMC result in the observed contact line pinning, receding contact angle measurements were performed using aqueous SDS solutions with concentrations of 0.05, 0.2, 0.5 and 1 CMC. Droplets were deposited on the hydrophobic substrate using the cannula needle and a syringe pump, as before. After deposition, the contact line was moved by first increasing the droplet volume to $2~\unit{\mu L}$. Second, by decreasing the volume by $0.8~\unit{\mu L}$ at a flow-rate of $0.1~\unit{\mu L/s}$ and third, by increasing the droplet volume again by $0.8~\unit{\mu L}$ also at a flow-rate of $0.1~\unit{\mu L/s}$. The droplets were imaged during their volume changes in order to measure the contact angle. The measurement procedure was repeated after 3, 6 and 10 minutes, in order to identify time dependent substrate wettability changes. The humidity inside the measurement chamber was set to $80\pm2\%$. At a RH of $80\pm2\%$, the evaporation was rather small so that the SDS concentration within the droplet can be considered constant. 

Figure~\ref{Fig5}A shows the four~receding contact angle measurements of the 0.05~CMC solution. The four~images overlaid in Figure~\ref{Fig5}B correspond to the moments in time indicated by the numbers in Figure~\ref{Fig5}A. Note that, independent of the time after droplet deposition, the contact line retracted smoothly. The contact line was also smoothly retracting during the initial receding contact angle measurement of the 0.2~CMC solution, see Figure~\ref{Fig5}C. However, contact angle jumps are clearly observed for the 0.2~CMC measurements repeated at 3, 6, and 10~minutes where the jumps become more apparent with the time after the first measurement. Also note, by comparing the droplet-air interface indicated by (3) and (4) in Figure~\ref{Fig5}D, that the droplet height increased after the sudden increase in contact angle (see movie in the Supporting Information). The observed contact angle jumps and corresponding changes in droplet height are a direct proof of  contact line pinning. Figures~\ref{Fig5}E and F show that the contact lines of the 0.5~CMC and the 1.0~CMC solution, respectively,  are again smoothly retracting. Thus Figure~\ref{Fig5} demonstrates that, indeed, at SDS bulk concentrations of approx. 0.2~CMC, the wettability of the PFDTS-coated surface is modified such that it results in contact line pinning of the SDS laden droplet.  

\subsection{SDS concentration dependent adsorption}

\begin{figure}[h] 
	\begin{center}          
		\includegraphics[width=1\columnwidth]{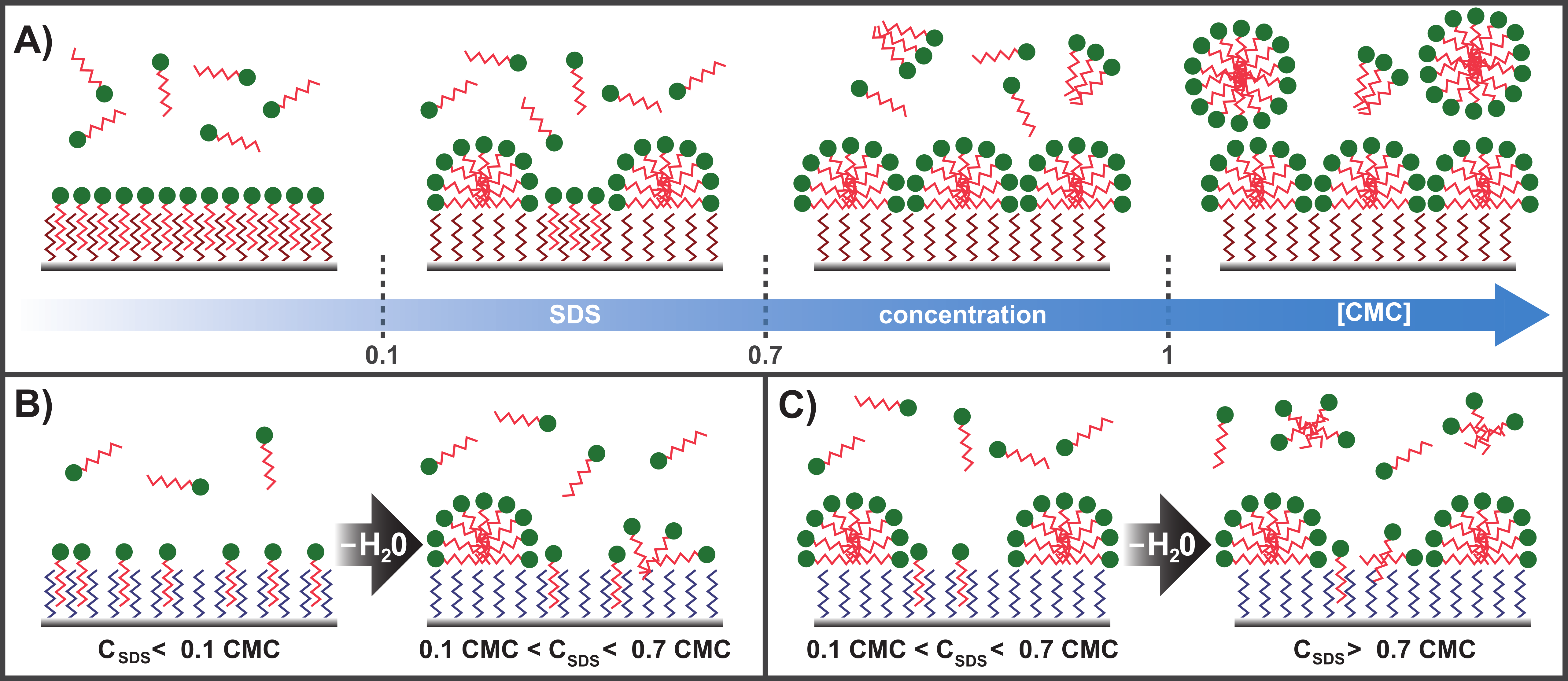} 
		\caption{A) Schematic representation of concentration dependent SDS adsorption based on the work of Duan~\emph{et al.}~\cite{SDSadsorption} B) Illustration of disordered SDS aggregate formation during the evaporation of SDS laden sessile droplets with initial concentrations below 0.1 CMC and (C) that for initial concentrations between 0.1 and 0.7 CMC.}
		\label{Fig6}
	\end{center}  
\end{figure}

The presented experimental results suggest that the local contact angle minimum is induced by a modification of the surface wettability due to adsorbed surfactant molecules. It has been shown that ionic surfactants with long aliphatic carbon chains like CTAB or SDS  adsorb to an amorphous hydrophobic substrate such as PFDTS coated silicon.~\cite{SurfOrganization} Experiments have been performed for static conditions by dipping the substrate in the surfactant solution. The structure of the adsorbed molecules has been shown to strongly depend on the surfactant concentration, see Fig.~\ref{Fig6}A. 
At concentrations below 0.12~CMC, surfactant molecules adsorb on the surface to form an amorphous layer, with the hydrophobic tail pointing towards the molecules coating the surface. At concentrations exceeding 0.12~CMC, ordered structures were found to form starting with the sparse formation of dome-shaped half-micelles that were found to fully cover the surface at concentrations above 0.73~CMC, see Fig.~\ref{Fig6}A. The formation of surface aggregates at concentrations below the CMC, i.e. at concentrations below the threshold for bulk aggregate formation, results from the hydrophobic surface that provides a nucleation site for aggregate formation.

Based on the introduced literature,~\cite{SurfOrganization} a potential mechanism for the observed contact line pinning is discussed. For an evaporating sessile droplet with $c_{i}<0.1~\unit{CMC}$, at first, the SDS molecules will form an amorphous layer on the substrate below the droplet. Later, as the surfactant concentration increases through evaporation, the threshold concentration for surface aggregate formation is reached. However, it is expected that the pre-adsorbed amorphous surfactant layer hampers the ordered formation of dome shaped micelles since energy is required to desorb the surfactant monolayer at the location of such a dome shaped micelle. Therefore, it is speculated that in the case of a dynamically changing surfactant concentration such as that found in an evaporating droplet, disordered surface aggregates are formed, see Figure~\ref{Fig6}B. 
A disordered layer of adsorbed surfactant aggregates will have non-uniform wetting properties through which it can induce contact line pinning. 
As the evaporation process continues, the surfactant concentration further increases resulting in a potential reorganization of the disordered surfactant layer into a homogeneous layer with corresponding homogeneous surface wetting properties. The homogenous wetting properties will result in a smoothly retracting contact line at a higher contact angle and an absence of contact line pinning, exactly as observed in Figures~1-3 after the appearance of the contact angle minimum. For the sessile droplets with a higher initial surfactant concentrations, i.e. with $0.1<c_{i}<0.7~\unit{CMC}$, the surface will be only partly covered with an amorphous layer in the first moments of the evaporation, Fig.~\ref{Fig6}C. Therefore, it is expected that $0.1<c_{i}<0.7~\unit{CMC}$ result in a more homogeneous surface wettability. Indeed, this is in perfect agreement with the results shown in Figure~\ref{Fig2}B, i.e.  the contact angle at the local minimum grows with increasing initial concentration of the SDS solution. In an attempt to image the shape of the surfactant structures in the range of low ($c_{SDS}<0.5~\unit{CMC}$) SDS concentrations, we used AFM imaging in liquid. Unfortunately, scans appeared blurry, probably due to the fact that the AFM tip disturbs and interacts with loosely packed surfactant molecules. A similar effect was reported previously.~\cite{SDS_AFM_1, SDS_AFM_2}

The receding contact angle measurements showed that contact line pinning increases with waiting time (Figure~\ref{Fig5}) which suggests that that disordered adsorption of surfactant molecules is enhanced at a stationary contact line. For the evaporating sessile droplets,  no dependence of the local contact angle minimum on the total evaporation time (Figure~\ref{Fig3}) was observed. This discrepancy is most likely a result of the non-stationary contact line of the evaporating SDS laden droplets in contrast to the stationary contact line during the receding contact angle measurements.
 
\section{Conclusions}

It has been found that the contact angle of an SDS laden droplet with a concentration below 0.5~CMC first decreases, then increases, and finally decreases resulting in a local contact angle minimum. The minimum contact angle was found to be substantially lower than the static receding contact angle and it decreased with decreasing initial SDS concentration. Furthermore, the bulk SDS concentration at the local contact angle minimum was found to decrease with a decrease in the initial SDS concentration. The location of the observed contact angle minimum relative to the normalized evaporation time and its minimum value proved to be independent of both the relative humidity and the droplet volume. The observed contact angle minimum has been attributed to contact line pinning induced by surfactant concentration dependent aggregate formation on the substrate below the droplet. The present work highlights the complexity of the evaporation sessile liquid-surfactant droplets and the influence of surfactant-substrate interactions on the evaporation process.

\section*{Acknowledgements}

This work is part of an Industrial Partnership Programme of the Foundation for Fundamental Research on Matter (FOM), which is financially supported by the Netherlands Organisation for Scientific Research (NWO). This research program is co-financed by Oc\'{e} Technologies B.V., University of Twente, and Eindhoven University of Technology.

Additionally, we would like to thank all members of the FIP program for useful discussions and all their suggestions. We would also like to thank Guillaume Lajoinie for help with the humidity control setup.

\section{Supporting Information}

AFM topographical image and the height profile of a PFDTS coated substrate; Movie illustrating droplet receding measurement

\bibliography{BibliographySDS}
	

\end{document}